\documentclass[prl,twocolumn,showpacs,floats]{revtex4}
\usepackage{epsfig}
\usepackage{graphicx}

\begin{document}

{\bf Reply to ``Comment on `The nature of slow dynamics in a minimal
model of frustration-limited domains'" by J. Schmalian, P.G. Wolynes
and S. Wu}. In a recent paper\cite{plg_drr} we reported simulation results for a
minimal model of frustration-limited domains with energy functional
\begin{eqnarray}
&&  
\hspace{-5mm}
\beta {\cal H}[\phi({\bf r})]  =  
\nonumber\\
&&
\int d{\bf r}
\left[\frac{1}{2}\phi({\bf r}) \left(\tau + k_{0}^{-2}(\nabla^2 +
k_0^2)^2\right) \phi({\bf r}) 
{\lambda \over 4!}\phi^4({\bf
r})\right].
\label{equ:hamiltonian}
\end{eqnarray}
We concluded that in all but a very small unexplored portion of the phase
diagram the Langevin equation corresponding to Eq.~{\ref{equ:hamiltonian},
\begin{equation}
{\partial \phi({\bf r}) \over {\partial t}} = -{\delta {\cal H}[\phi]
\over{\delta \phi({\bf r})}} +\eta({\bf r},t)
\label{equ:langevin}
\end{equation}
with Gaussian random force $\eta({\bf r},t)$,
does not generate glassy dynamics. 
Although relaxation times do grow
dramatically as a function of inverse temperature, we found no
evidence for non-exponential multi-step relaxation.
A simple Hartree approximation is in fact sufficient to capture the
computed dynamics almost quantitatively.  The
validity of a harmonic reference system, we reasoned, demonstrates that activated
processes and the possibility of an entropy crisis are not essential
for dynamics at temperatures above (but not extremely close to) the
fluctuation-induced first order transition temperature, $\tau_{\rm tr}$.
We further noted that facile growth of long-range order upon cooling below
$\tau_{\rm tr}$ renders the study of a stationary, supercooled disordered state
impossible.
The relevance of our conclusions for the stripe glass scenario of
Schmalian and Wolynes has recently been questioned by Schmalian {\em
et al.}\cite{schmalian_comment}.  In particular, they point out that our simulations span
a range of effective temperatures above that at which glassiness sets in, 
$T_{\rm c}$, as predicted by either the self-consistent screening approximation (SCSA)
or dynamical mean-field theory (DMFT).

First, we discuss the nature of the approximations used in \cite{plg_drr} and
\cite{schmalian_comment} to calculate $T_{\rm c}$.  
Schmalian {\em et al.} describe
SCSA and DMFT as more sophisticated approaches than the
mode-coupling approximation (MCA) analyzed in our work.  
Regardless of the level of sophistication of these approximations, we
chose to focus on the MCA because it is the clear analog of 
G\"otze's mode coupling theory (MCT) for relaxation of density 
fluctuations in liquids\cite{goetze}.
Although MCT overestimates $T_{\rm c}$
in model systems\cite{nauroth}, simulations show a robust correlation
between the predicted ergodic-nonergodic transition and
the onset of caging in supercooled liquids\cite{brumer,sastry}.   We found
no hint of caging behavior in our simulations, even 
below the values of $\tau$ for which the MCA predicts loss of ergodicity.  
Furthermore, the correspondence between static correlations and 
dynamical relaxation in our minimal model is much stronger than that exhibited
by supercooled liquids, and the non-Arrhenius temperature dependence of 
correlation times is coupled to exponential, rather than multi-step, relaxation.
These prominent features of the dynamics generated by (\ref{equ:langevin}), while
interesting, are therefore not generically similar to those found in supercooled
liquids.

\begin{figure}
\includegraphics[width=0.8\linewidth]{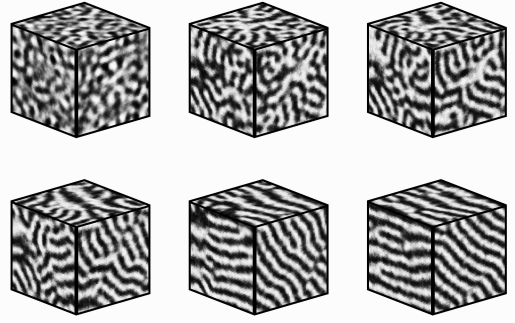}
\caption{Coarsening of a configuration whose
structural correlations are consistent with the disordered solution of DMFT 
at $\tau=-0.2$, $q_{0}=0.5$, and $\lambda=1$. 
Field values $\phi({\bf r})$ at the surface of our (periodically replicated) 
simulation cell are shown for times
$t=0$ (top left), $t=500$ (top middle), $t=1000$ (top right),
$t=2000$ (bottom left), $t=5000$ (bottom middle), and $t=10,000$ (bottom right).}
\end{figure}

In the conclusion of our work we mentioned that growth of long-range
lamellar order occurs even slightly below $\tau_{\rm tr}$, indicating a lack
of stable glassy behavior at temperatures below those studied thoroughly
in \cite{plg_drr}.  Here, we address the criticism of Schmalian {\em et al.}
by presenting detailed results for $\tau < \tau_{\rm tr}$.
For several disordered configurations representative of the
equilibrium state at $\tau=-0.14$, $q_{0}=0.5$ and $\lambda=1$,
we have performed instantaneous quenches to
$\tau=-0.2$, the temperature at which 
Schmalian {\em et al.} predict the onset of glassiness, and have followed the subsequent evolution in time.  
Instead of the aging behavior
typical of a supercooled liquid \cite{kob} (weak time
dependence of one-time quantities and relative {\em stability} of the
disordered phase), coarsening occurs, i.e., the
modulated order of the ground state rapidly appears in all regions of our simulation 
cell, and ordered domains gradually align.  
As a more explicit test of the stability of the glassy state described in 
\cite{schmalian_comment}, we have constructed initial configurations representative of the DMFT solution of Schmalian {\em et al.} for $\tau=-0.2$, drawing values
for the Fourier components of $\phi({\bf r})$ from appropriate Gaussian distributions\cite{deem}.
A typical trajectory originating at such a glassy state is depicted in Fig.~1.
Again, the system steadily coarsens towards the
ordered state.  In Fig.~2 we show the peak of the structure factor, $S_0(t)$, as a function of time for examples of both types of initial conditions.
After a short transient, the two results merge and 
grow approximately with the power law expected from previous
studies of models in the Brazovskii class\cite{elder}.
These low-temperature results suggest that the topography of
(\ref{equ:hamiltonian}) in this region of configuration space differs from
that envisioned by Schmalian, Wolynes and coworkers and that standard
entropy crisis arguments must be modified in this context.  
In particular, the aging dynamics we have computed contrast strongly with the aging of a supercooled liquid, implying that the relevant region of the landscape
(\ref{equ:hamiltonian}) does not closely resemble
that of a supercooled liquid at low temperatures\cite{kob,crisanti}.

\begin{figure}
\includegraphics[width=0.8\linewidth]{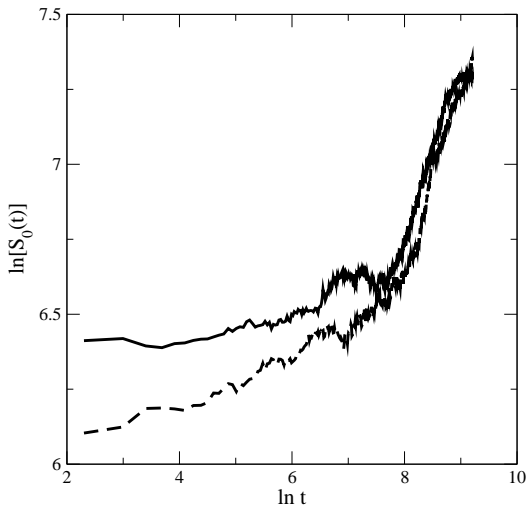}
\caption{Aging of the peak of the structure factor, $S_0(t)$, for two trajectories 
at $\tau=-0.2$, $q_{0}=0.5$, and $\lambda=1$.  
The dashed line corresponds to an initial configuration representative of equilibrium at $\tau=-0.14$. 
The solid line corresponds to an initial configuration consistent with the statistics of 
DMFT described in \cite{schmalian_comment}.  
Although these results are anecdotal, late stage growth is consistent with
a power law $S_0(t)\propto t^\alpha$ with exponent $\alpha \approx 0.5$.
}
\end{figure}

We do not deny that experiments in high temperature superconducting
materials reveal slow wandering of stripes.  This behavior may indeed
be consistent with coarsening in two dimensions, which permits very
slow rearrangement of modulated structures.  In addition disorder and
other factors neglected in (\ref{equ:hamiltonian}) may also 
stabilize the quasi-disordered
dynamics of hole-rich and hole-poor regions, perhaps giving rise to
glassy dynamics rather than simple coarsening.  One possibility is
that more complicated terms should be included in (\ref{equ:hamiltonian}) that would
stabilize the glassy state\cite{ferro_glass}.  Another possibility is that a stable
glass state arises only in the limit
of strong coupling ($\lambda \rightarrow \infty$).  Investigation of these possibilities
is certainly a worthy direction for future research.{~\\}{~\\}{~\\}

{\noindent Phillip L. Geissler~\\~{\em Department of Chemistry}~\\~{\em University of California, Berkeley}~\\~{\em Berkeley, California 94720}{~\\}{~\\}}

{\noindent David R. Reichman~\\~{\em Department of Chemistry and Chemical Biology}~\\~{\em Harvard University}~\\~{\em Cambridge, Massachusetts, 02138}}

\renewcommand\refname{}

\end{document}